\documentclass[pra,aps,twocolumn,10pt,showpacs,groupedaddress,superscriptaddress,floatfix,notitlepage]{revtex4-2}
\usepackage{amsmath,amsfonts,amssymb,graphics,graphicx,epsfig,color,times}%,bbm}
\usepackage[utf8x]{inputenc}
\usepackage{color}
\usepackage{bbm, dsfont} 
\usepackage{subfigure}
\usepackage{hyperref}
\usepackage{mathrsfs}
\usepackage{verbatim}

\usepackage{graphicx}% Include figure files
\usepackage{dcolumn}% Align table columns on decimal point
\usepackage{bm}% bold math
\usepackage{xcolor}
\begin{document}

%\preprint{APS/123-QED}

\title{\bf A photonic engine fueled by quantum-correlated atoms}% Force line breaks with \\

\author{Chimdessa Gashu Feyisa}
 \affiliation{Molecular Science and Technology Program, Taiwan International Graduate Program, Academia Sinica, Taiwan}
\affiliation{Department of Physics, National Central University, Taoyuan City 320317, Taiwan}
 \affiliation{Institute of Atomic and Molecular Sciences, Academia Sinica, Taipei 10617, Taiwan}

\author{H. H. Jen}
\email{sappyjen@gmail.com}
%\author{\textsuperscript{1,2}}
%\author{author4\textsuperscript{1}}
 \affiliation{Molecular Science and Technology Program, Taiwan International Graduate Program, Academia Sinica, Taiwan}
 \affiliation{Institute of Atomic and Molecular Sciences, Academia Sinica, Taipei 10617, Taiwan}
\affiliation{Physics Division, National Center for Theoretical Sciences, Taipei 10617, Taiwan}

\date{\today}% It is always \today, today,

\begin{abstract}
Entangled states are an important resource for quantum information processing and for the fundamental understanding of quantum physics. An intriguing open question would be whether entanglement can improve the performance of quantum heat engines in particular. One of the promising platforms to address this question is to use entangled atoms as a non-thermal bath for cavity photons, where the cavity mirror serves as a piston of the engine. Here we theoretically investigate a photonic quantum engine operating under an effective reservoir consisting of quantum-correlated pairs of atoms. We find that maximally entangled Bell states alone do not help extract useful work from the reservoir unless some extra populations in the excited states or ground states are taken into account. Furthermore, high efficiency and work output are shown for the non-maximally entangled superradiant state, while negligible for the subradiant state due to lack of emitted photons inside the cavity. Our results provide insights in the role of quantum-correlated atoms in a photonic engine and present new opportunities in designing a better quantum heat engine.
\end{abstract}

%\keywords{Photonic engine, Efficiency, Cavity photons, Quantum correlated atoms, Effective reservoir}%Use showkeys class option if keyword
\maketitle

%\tableofcontents

\section{\label{sec: I} Introduction}

Heat engines, a special type of thermal machines, have made great strides in the industrial revolutions. 
Steam engines have revolutionized textile technology and plant engineering, while Otto and diesel heat engines have prevailed in electric motors \cite{ar1}. Microwave amplification by stimulated emission of radiation (maser), often regarded as the birth of quantum thermodynamics, functions on the basis of Otto engine cycles \cite{ar2}. Despite many advances, the efficiency of any heat engine, no matter how modern, is limited by Carnot efficiency - an ideal efficiency for a fully reversible and frictionless Carnot cycle that does not lead to entropy change. The Carnot efficiency thus posits a universal limit on the efficiency of any classical heat engine \cite{ar3, ar4}. Interest in optimizing the efficiency as well as output power has led to the development of the quantum heat engines, which are the integral parts of quantum thermodynamics \cite{ar4, ar5, ar6, ar7}. 

In this regard, the design of heat engines operating at the nano and atomic scales has advanced with the current era of technology \cite{ar8,ar9, ar10, ar11, ar12}, which resonates with Feynman's statement that there is plenty of room at the bottom, and various theoretical and experimental models have been reported \cite{ar13, ar14, ar15, ar16}.  Such quantum thermal devices can provide a paradigm to combine quantum mechanics and thermodynamics, apart from their promising advantages for the use of industrial applications. One of the interesting initial steps was a single atom heat engine realized experimentally \cite{ar16} and extended into other platforms involving quantum coherences \cite{in1,in2,in3,in4,ar17} and collective phenomena \cite{co1,co2,co3}. It has been shown that the maximum atomic coherence dubbed in a single atomic beam can boost the efficiency of a photonic engine to nearly unity and increases the work output quadratically with the mean number of single atoms \cite{ar17}. Here a two-level atom has served as a non-thermal reservoir for the cavity radiation, driving it to a finite and positive temperature at steady state. It is a well known fact that a state with quantum coherence is at the boundary of classical and quantum states. Therefore, a full quantum description can go beyond a single particle scheme and should involve entangled states, whose effects are worth investigations. Superradiant \cite{ar44} and subradiant states are examples of maximally entangled states with symmetric and antisymmetric combinations of the dipole moments of two atoms. Other interesting two-atom entangled states are superpositions of their doubly excited and common ground states \cite{ar18}. In general, entangled states are a key for the basic understanding of quantum physics and a resource for many interesting phenomena in quantum information, computing, sensing, communication among others \cite{ar19,ar20,ar21}. An intriguing open question would be whether entanglement plays some role in efficient work extraction processes in quantum heat engines \cite{en1,en2,en3,en4,en5}. 

In this work, we investigate whether it is possible to achieve the maximum efficiency recorded in Ref. \cite{ar17}, where barium atoms in an optical cavity are used with $^1S_0$ to $^3P_1$ atomic transition at $791$ nm, and increase the work output in a quantum heat machine with quantum entanglement. We considered a beam composed of pairs of two-level atoms that exhibit quantum entanglement, which has been shown to provide a wider temperature range for the cavity field than single-atom heat engines \cite{ar22,ar23}. Two-qubit entanglement is one of the most studied topics in various trapping and cooling systems \cite{ar24,ar25,ar26}. Symmetric and anti-symmetric superpositions of two atoms can be prepared using a coherent laser drive, while two-photon entangled states can be created via Raman excitation processes \cite{ar27}. Several natural systems such as electron-positron pairs exhibit these entangled states \cite{ar28}, which are readily controlled by atom-atom collisions or dissociation of molecules in a cavity \cite{ar27}. We assume that the entangled atoms pass through a cavity field, which is a working medium for the quantum heat engine. We aim to clarify the best operating parameters regime of the quantum heat engine, taking into account the two key performance indicators, efficiency and work output. We have shown that the cavity field produces more work with higher efficiency when excited by the atomic beams with finite quantum correlations, whereas maximally entangled Bell states forbid an effective work extraction process. In addition, phase-controlled energy transfer from the quantum correlated atoms to the field is observed due to the symmetrical nature of the Jaynes-Cumming model. In our photonic engine, the effective reservoir has supplied both heat and work to the engine, unlike the uncorrelated reservoir which can only provide heat energy. In this case, high engine efficiency beyond a classical thermodynamic limit is not surprising. Our results push forward experimental works on quantum thermal machines driven by the non-thermal reservoir that incorporates all types of quantum correlations and can be extended to other platforms with large degrees of freedom.

The rest of the paper is organized as follows. We first introduce the scheme of the photonic engine and Hamiltonian of the system in Sec. \ref{sec: II}. In Sec. \ref{sec: III}, we derive the equation of motion for the cavity photons. We then introduce a photonic engine cycle and derive the relevant thermodynamic quantities in Sec. \ref{sec: IV}. In addition, we present the engine efficiency and the work output with detailed study of the influence of finite quantum entanglement. Finally, we provide concluding remarks in Sec. \ref{sec: V}.
\section{\label{sec: II} Hamiltonian of the system}
\begin{figure}
\begin{center}
\resizebox{0.5\textwidth}{!}{%
\includegraphics{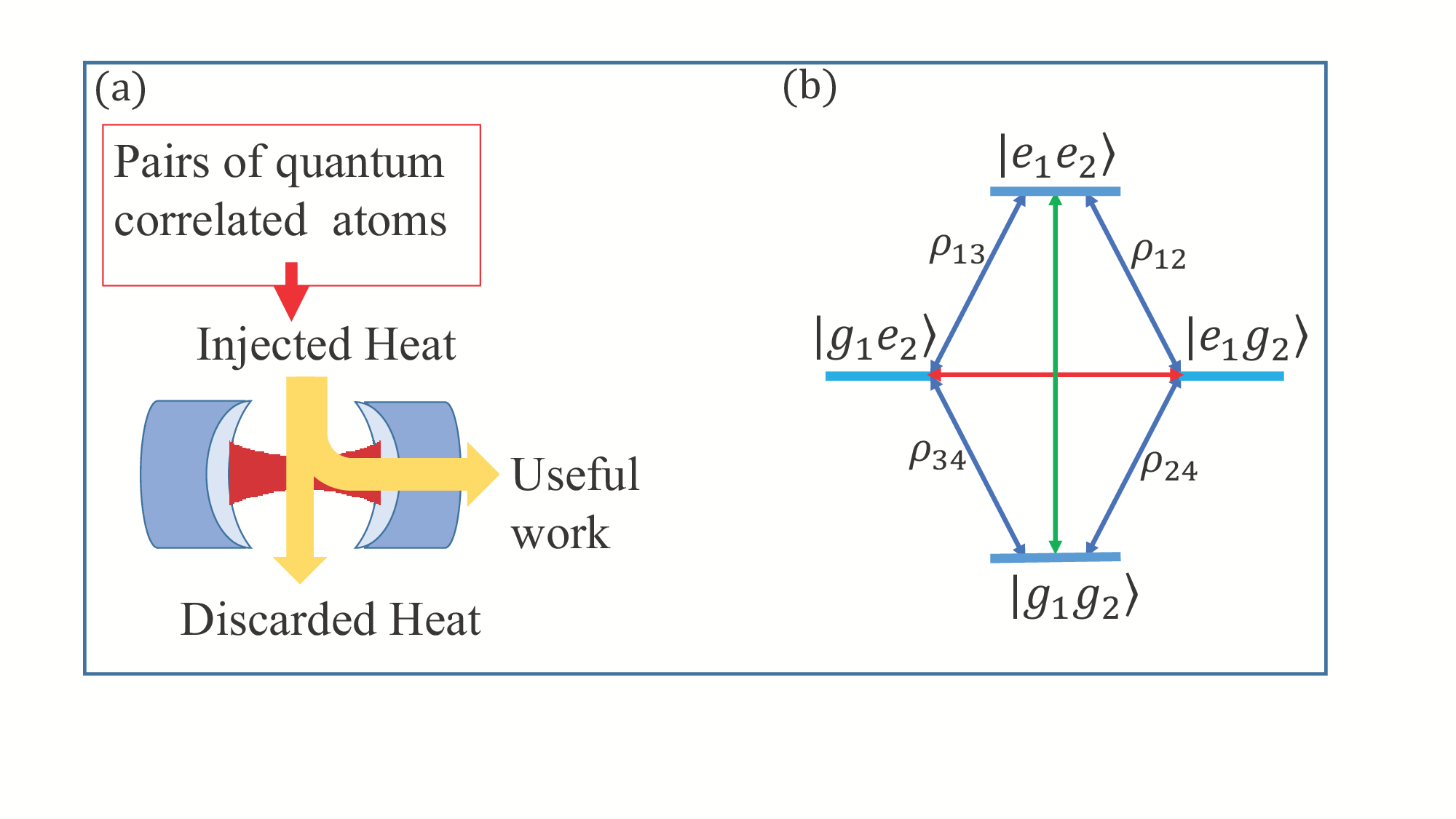}}
\caption {Schematic of the photonic engine: (a) The cavity field interacts with pairs of two-level atoms acting as hot and cold non-thermal baths based on its states. The cyclic processes of the interaction is able to be controlled by the relevant system parameters. (b) Atomic coherences that lead to quadratic growth of the radiation pressure with the mean number of atom pairs.}\label{f1}
\end{center}
\end{figure}

The model system consists of two two-level atoms that interact with a quantized cavity field as shown in Figs. \ref{f1}(a) and \ref{f1}(b). The free and interaction Hamiltonians, $H_s$ and $H_I$ respectively, for the system can be written in the form \cite{ar30,ar31}:
\begin{eqnarray}
 \hat H_s&=&\hat H_c+\hat H_a={\omega_c}\hat a^\dag\hat a+\sum^2_{i=1}\frac{\omega_i}{2}\hat\sigma^+_i\hat\sigma^{-}_i,\\
 \hat H_I&=&\sum^2_{i=1}\left[\frac{\delta_i}{2}\hat\sigma^{z}_{i}+g\left(\hat{a}\hat\sigma^{+}_i+\hat\sigma^{-}_i\hat { a}^\dag\right)\right]\label{2},
\end{eqnarray}
where $\hat H_c$ and $\hat H_a$ are the unperturbed Hamiltonians of the cavity field and a pair of two-level atoms. Here $\hat a$ represents the annihilation operator of the cavity field with a frequency mode $\omega_c$, $g$ denotes the atom-field coupling constant, $\delta_i=\omega_i-\omega_c$ are the detuning frequencies between the cavity field and the atoms, where $\omega_{i}$ are the transition frequencies of each atoms whose raising and inversion operators are represented by $\hat\sigma^{+}_ {i}=(\hat\sigma^{-}_{i})^\dag=|e_i\rangle\langle g_i|$ and $\hat\sigma^{z}_{i}=|e_i\rangle\langle e_i|-|g_i\rangle\langle g_i|$, in which $|g_i\rangle$ and $|e_i\rangle$ are ground and excited states of $i$-th atoms. Pairs of quantum correlated atoms in Fig. \ref{f1}(b) are continuously injected into the cavity one by one, changing the state of the cavity field due to the light-atom interaction in the cavity system. The mean photon number of the cavity field changes due to expansion and compression processes that would occur when the externally prepared quantum correlated atoms pass through a cavity. Therefore, the engine generates work output by applying radiation pressure on the output cavity mirror. More details about the engine cycle can be found in Sec. \ref{sec: IV}. 

Interaction of the cavity field with a single two-level atom has been considered in Ref. \cite{ar17}, in which each atom is externally prepared in a coherent superposition of the ground and excited states by a laser pump after passing through a nano-hole array aperture and before interacting with a cavity field. Repeated interactions of the excited atoms with a cavity field in a short time interval have resulted in high work output and engine efficiency exceeding a classical Carnot limit. For a single two-level atom interacting with a cavity field, only atomic coherence is present as a quantum feature. However, if we consider even the simplest many-particle system - a pair of quantum correlated atoms and its interaction with a cavity field described in Eq. \ref{2}, novel quantum features can emerge from the designated atom-atom correlations. This opens up a space to explore whether these quantum correlations are a useful resource for engine operations.
\section{\label{sec: III} Equation of motion for cavity photons}
The interaction of atom with the cavity field could be described by a micro-maser type master equation \cite{ar32}. 
We assume that pairs of two-level atoms are continuously injected into the cavity, where they interact with the cavity field for a time $\tau$ before exiting the cavity system. We explore scenarios in which atoms traverse the cavity in much less time than the characteristic lifetime $1/\kappa$ of the cavity field while $\kappa$ is cavity field decay rate. Therefore, the atom-cavity system undergoes a unitary evolution during this short interaction time \cite{ar31}. The initial density operator of the system before the interaction can be written as $\rho_{ac}(t_i)=\rho_a\otimes\rho_c(t_i)$, where $\rho_{c}(t_{i})$ is the density operator of the cavity field prior to the arrival of atoms and $\rho_a$ is the 
initial density operator of a single pair of two-level atoms. The density operator of the cavity field evolves after interacting with the atoms in a time $\tau$ and can be put in the form \cite{ar32,ar33,ar34}
\begin{eqnarray}
\rho(t_i+\tau)={\rm Tr}_a[U(\tau)\rho_{ac}(t_{i})U^\dag(\tau)],\label{15.0}
\end{eqnarray}
where $U(\tau)=\exp(-i\hat H_I\tau)$ is the unitary time evolution operator of $\hat H_I$, and ${\rm Tr}_a$ represents a trace operation taken to remove the atomic variables.

We next introduce the general initial state of atomic pairs in a matrix form, $\rho_a$, which can be written in 
the standard basis of the two atoms $\{|e_1e_2\rangle,|e_1g_2\rangle,|g_1e_2\rangle,|g_1g_2\rangle\}$ as:
\begin{eqnarray}
 \rho_a=\begin{pmatrix}
    \rho_{11}&\rho_{12}&\rho_{13}&\rho_{14}\\
    \rho_{21}&\rho_{22}&\rho_{23}&\rho_{24}\\
    \rho_{31}&\rho_{32}&\rho_{33}&\rho_{34}\\
    \rho_{41}&\rho_{42}&\rho_{43}&\rho_{44}\\
   \end{pmatrix},\label{15}
\end{eqnarray}
where the non-diagonal (diagonal) elements $\rho_{ij,i\neq j} (\rho_{ij,i=j})$ represent the atomic coherences (populations). The state described in Eq. (\ref{15}) contains various quantum states in which the amount of entanglement between the atoms can be calculated by concurrence formula \cite{ar35,ar36}: $C=\text{max}[0,\Lambda],$  where $\Lambda=\sqrt{\lambda_1}-\sqrt{\lambda_2}-\sqrt{\lambda_3}-\sqrt{\lambda_4}$, and $\lambda_i$ are eigenvalues of a Hermitian operator $\hat m=\sqrt{\sqrt{\rho_a}\tilde{\rho}_a\sqrt{\rho_a}}$ with $\tilde{\rho}_a=(\sigma_y\otimes\sigma_y)\rho^\ast_a(\sigma_y\otimes\sigma_y)$ being the complex conjugation of $\rho_a$ while $\sigma_y=\begin{pmatrix}0&-i\\i&0\\\end{pmatrix}$ is the Pauli matrix. Concurrence is zero for product states and unity for maximally entangled state as for the case of well known EPR states \cite{ar35,ar36,ar37}. 
                                                                                                                                                                                                                                                                
We solve the master equation for the cavity photons explicitly using Eqs. (\ref{15.0}) and (\ref{15}). Pairs of quantum correlated atoms and a vacuum environment are combined to form an effective reservoir engineered for the cavity field. The equation of motion for the cavity field under this effective reservoir can be obtained as (see Appendix \ref{apa} in more detail),
\begin{eqnarray}
\frac{d\rho(t)}{dt}&=&-i[H_{\rm eff},\rho(t)]+\gamma\bar n_{th}[2\hat a^\dag\rho(t)\hat a-(\hat a\hat a^\dag\rho(t)\nonumber\\&+&\rho\hat a\hat a^\dag)]+\gamma(\bar n_{th}+1)[2\hat a\rho(t)\hat a^\dag-(a^\dag\hat a\hat\rho+\rho\hat a^\dag\hat a)]\nonumber\\&+&\mu(2\hat a\rho(t)\hat a-\hat a^{2}\rho(t)-\rho(t)\hat a^{2})+\mu^\ast(2\hat a^\dag\rho(t)\hat a^\dag\nonumber\\&-&\hat a^{\dag2}\rho(t)-\rho(t)\hat a^{\dag2}).\label{cm2}
\end{eqnarray}
In this equation, the first term illustrates a uniform evolution of the cavity field after tracing out the atomic variables, the second and third terms represent Lindblad dissipations, and the last two terms show two-photon de-excitations and excitations, which can be ignored when single two-level atoms pass through a cavity. The coefficients of the Lindblad dissipations: $\bar n_{th}=\frac{p_1}{\kappa/2+(p_2-p_1)}$ and $\gamma=\kappa/2+(p_2-p_1)$ indicate the thermal photon number and an effective decay rate of the cavity photons due to the cavity attenuation rate $\kappa$ and atomic properties. Other quantities involved in Eq. (\ref{cm2}) are given by
\begin{eqnarray*}
p_1&=&g^2\tau N_{pair}\text{Re}(R)[\rho_{11}+\rho_{22}/2+\rho_{33}/2+\rho_{23}+\rho_{32}],\\
p_2&=&g^2\tau N_{pair}\text{Re}(R)[\rho_{44}+\rho_{22}/2+\rho_{33}/2+\rho_{32}+\rho_{23}],\\
\mu&=&g^2\tau N_{pair}\rho_{41}\text{Re}(R)e^{i\delta\tau},
\end{eqnarray*}
with $N_{pair}=r_a\tau$ defining an effective number of atomic pairs in the cavity, $R=e^{i\delta\tau/2}\big[\text{sinc}(\frac{\delta\tau}{2})+\frac{2i}{\delta\tau}\big(\text{cos}(\frac{\delta\tau}{2})-\text{sinc}(\frac{\delta\tau}{2})\big)\big],$ and $\delta_1=\delta_2=\delta$. The effective Hamiltonian $H_{\rm eff}$ is
\begin{eqnarray}
H_{\rm eff}&=&gr_a\tau(\alpha\hat a^\dag+\alpha^\ast\hat a)-r_a(\rho_{44}-\rho_{11})\text{Im}(R)(g\tau)^2\hat a\hat a^\dag\nonumber\\&-&\frac{r_a(\rho_{22}-\rho_{33})}{2}\text{Im}(R)(g\tau)^2,
\end{eqnarray}
where $\alpha=(\rho_{13}+\rho_{24}+\rho_{12}+\rho_{34})\text{sinc}(\frac{\delta\tau}{2})e^{-i\delta\tau/2}$ denotes a parameter that drives the cavity field. The effective Hamiltonian $H_{\rm eff}$ governs the dynamics of the cavity photons in the presence of dissipation and squeezing terms triggered by the quantum correlated atoms. The second and third terms of $H_{\rm eff}$ characterize frequency pushing and pulling effects with a negligible contribution to the dynamics of the system. The master equation in Eq. (\ref{cm2}) describes the dynamics of the cavity field in a short interaction time $\tau$ with the pairs of atoms, which significantly depend on the atomic initial conditions. This formulation represents an extension to the single-atom picture when quantum correlations between atoms are neglected.  

We then obtain the steady-state solution of the mean number of cavity photons as (see Appendix \ref{apa2} for more detail):
\begin{eqnarray}
\bar n_{ss}=\langle\hat a^\dag\hat a\rangle_{ss}=\bar n_{th}+\frac{(gN_{pair})^2}{\gamma^2}|\alpha|^2.\label{mp4}
\end{eqnarray}
The cavity field involves two contributions: The thermal photon number from the effective reservoir and the number of photons from the quantum correlated atoms, respectively. We note here that the non-diagonal density matrix elements responsible for the entangled reservoir are included only in the first term of Eq. (\ref{mp4}). The term related to $|\alpha|$ involves only the initial atomic coherences for individual atom pairs that drive the cavity field. It is important to note that this additional term grows proportional to the square of mean number of atom pairs $N_{pair}=r_a\tau$, which leads to a phenomenon analogous to superradiance during the iso-energetic expansion of the cavity field and vanishes for $\alpha=0$ (more details in \ref{sec: IV}). The four maximally entangled Bell states do not contribute to this term since the relevant correlations from Bell states $\rho_{23}, \rho_{32}, \rho_{14}, \rho_{41}$ do not appear in $\alpha$. Therefore, the mean number of cavity photons is equal to the thermal photon number when the driving term is switched off, and we will show that a photonic engine never extract work output in this case. In other words, the effective reservoir acts as a passive state \cite{ar8} to the cavity field when the atoms are initially in maximally entangled Bell states.
\section{\label{sec: IV} Photonic engine cycle and efficiency}
Next, we analyze the efficiency of a photonic engine to extract work from a single effective reservoir. This purely quantum mechanical approach has been proposed theoretically and verified in experiments \cite{ar17,ar38,ar39}. The cavity photons do work due to the driving term in the effective Hamiltonian and exchange heat with the reservoir through the coherences during the dissipations process. The cavity photons as the engine's working fluid can run between different states of the reservoir that acts as a hot reservoir for $\alpha\neq0$ and cold reservoir for $\alpha=0$ analogous to an independent two-atom case \cite{ar17}. The efficiency of the engine can now exceed the Carnot limit \cite{ar39} and can be controlled by appropriate system parameters. Carnot efficiency limit is derived from the second law of thermodynamics which is restricted to the transfer of energy by heat alone, although energy transfer in the form of work can also occur without violating the laws of thermodynamics \cite{ar39,ar40}. In this case, quantum efficiency could increase by investing some energy to create the non-thermal reservoir. This fact is indicated by the driving term $\alpha$. However, improving efficiency by using the quantum advantage of the non-thermal reservoir is far cheaper than running the heat engine \cite{ar40}.
\subsection{Engine cycle} 
\begin{figure}
\begin{center}
\hspace{-2cm}{(a)\hspace{5cm}(b)}\\
\resizebox{.5\textwidth}{!}{%
\includegraphics{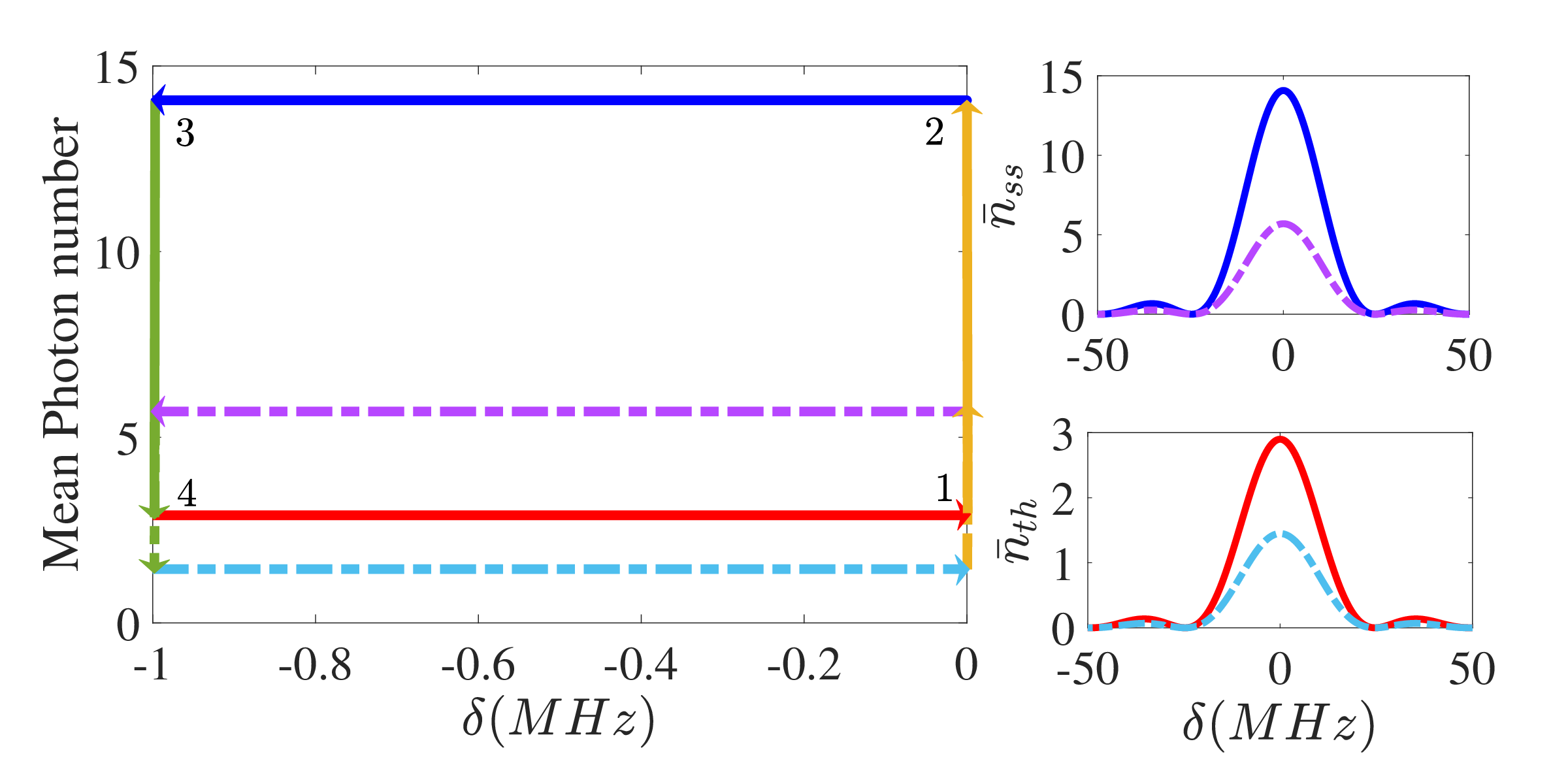}}
\caption{Photonic engine cycle and operations. (a) A photonic engine cycle analogous to the classical string cycle for $a=1, b=5, c=1, \phi=0, \kappa/2\pi=74\text{kHz}$, $g\tau=0.17$, and $N_{pair}=1$ (dash-dot curves) and $N_{pair}=2$ (solid curves). In the first stage (1$\rightarrow$2), the cavity field is heated by sending quantum-correlated atoms into the cavity. Hence, energy transfer from the effective reservoir to the cavity photons increases the internal energy of the engine. The heating phase is followed by an expansion (2$\rightarrow$3) at constant internal energy in response to the passage of quantum-correlated atoms through the cavity. The expansion can be controlled by tuning a frequency of the cavity photons $\omega_c$ with a fixed atomic transition frequency $\omega_a=\frac{\omega_1+\omega_2}{2}$. The next cycle is a cooling phase (3$\rightarrow$4) in which the engine releases its energy to the cold reservoir. The engine cycle is completed with a compression stage (4$\rightarrow$1), where the detuning frequency returns to its original value. (b) Average photon numbers of the cavity field becomes $\bar{n}_{th}$ (bottom panel) and $\bar{n}_{ss}$ (top panel), respectively, after interacting with an effective cold reservoir ($\alpha=0$) and an effective hot reservoir ($\alpha\neq0$).}\label{f2}
\end{center}
\end{figure}

We consider a similar photonic engine cycle as that in Ref. \cite{ar17} with some exceptions: Entanglement is incorporated to clarify the roles of quantum-correlated atoms, and we initialize our engine cycle with zero detuning where the system operates more efficiently. In the first step (1$\rightarrow$2), pairs of entangled atoms, described by a state $|\psi\rangle={\cal N}\big(a|e_1e_2\rangle+b\big(|g_1e_2\rangle+e^{i\phi}|e_1g_2\rangle\big)+c|g_1g_2\rangle\big)$ with a normalization constant ${\cal N}$, are sent into the cavity. They then impart their energy to the cavity field through resonant interactions with a quantized field confined within the leaky cavity walls. As a result, the internal energy of the engine increases rapidly and the cavity field undergoes iso-energetic expansion, which is the second stage (2$\rightarrow$3) of the engine cycle shown in Fig. \ref{f2}. Here the atom pairs and cavity field interact non-resonantly with one another. This effect lowers the internal energy of the cavity field as the engine does work during the expansion using the provided heat energy from pairs of quantum-correlated atoms. The remaining energy is discarded to the cold bath in the third engine cycle - isochoric cooling stage (3$\rightarrow$4). The cold bath could be described as discarded atomic beams after passing through the cavity \cite{ar38,ar400}. The engine cycle is closed by an iso-energetic compression (4$\rightarrow$1). In this step, the detuning frequency returns to its original value (resonance case) to begin the next cycle with the isochoric heating step. It is worth noting that the expansion and compression steps produce work by adjusting the cavity-field frequency, which is inversely proportional to the cavity volume \cite{ar17,ar400}. On the other hand, during the heating and cooling cycles, the engine does no work as the energy gap is fixed due to the constant frequency of the cavity field \cite{ar401}. However, there is heat exchange between the effective reservoir and the cavity field in these processes.
\subsection{Engine efficiency} 
The effective cavity field temperature during iso-energetic processes is given by a thermodynamic relation \cite{ar401, ar41} $T_{\rm eff}=dQ/dS.$ In this relation $S$ is the entropy of a single-mode field in the steady state. The entropy of the stationary cavity field during expansion and compression processes is the same and varies with thermal photon number as \cite{ar6,ar16,ar17,ar42}: $S/k_B=(\bar{n}_{th}+1)\text{log}(\bar{n}_{th}+1)-\bar{n}_{th}\text{log}\bar{n}_{th}$. This suggests that the entropy change is vanishing during the heating and cooling phases, since $\bar{n}_{th}$ is constant in both cases. The infinitesimal change in internal energy of the engine can be expressed as $dU=\omega_cd\bar n_{i}+\bar n_{i}d\omega_c$, where $\bar{n}_{i}=\bar{n}_{ss}$ and $\bar{n}_{i}=\bar{n}_{th}$ for the expansion and compression steps, respectively. In this expression, $\omega_cd\bar n_{i}$ is defined as heat exchange between system and reservoir since the energy exchange is triggered by the internal reconfiguration of the quantum system \cite{ar401,ar42}. On the other hand, the second term $\bar n_{i}d\omega_c$ indicates energy transfer by performing work, which can be controlled by tuning the cavity frequency \cite{ar39,ar42}. 
The effective temperature of the cavity field during the expansion stage can be written in the form
\begin{eqnarray}
T_{\rm eff}=\frac{dQ}{dS}=\frac{\omega_c\bar n_{ss}}{\bar{n}_{th}k_B\text{log}(1+\frac{1}{\bar{n}_{th}})},
\end{eqnarray}
where we have used $\frac{d\omega_c}{d\bar n_{th}}=-\frac{\omega_c}{\bar n_{th}}$ and $dQ=-\bar n_{ss}d\omega_c$ due to the internal energy constraint ($dU=0$) of iso-energetic processes. The same procedure can be followed to find the effective temperature of the cavity photons during the compression process.  

The net work done and heat exchanged during engine operation are given by $W_{net}=W_{12}+W_{23}+W_{34}+W_{41}$, $Q_{inj}=Q_{12}+Q_{23}+Q_{34}$, and $Q_{dis}=Q_{41},$ where $Q_{inj}$ and $Q_{dis}$ represent the injected and discarded heat. We note that $Q_{12}=-Q_{34}$ since the heat injected into the cavity by a hot reservoir during the heating stage returns to the cold reservoir during the cooling stage \cite{ar17}. Moreover, the heat input due to the thermodynamic entropy change during expansion is canceled by heat dissipated during compression. Thus, $Q_{inj}=Q_{23}$ and $W_{net}=W_{23}+W_{41}=Q_{23}-Q_{41}$. Considering the definitions of work and heat, and referring to the engine cycle, the efficiency of the quantum heat engine is given by \cite{ar17}:
\begin{eqnarray}
 \eta&=&\frac{W_{net}}{Q_{inj}}=1-\frac{Q_{41}}{Q_{23}}=\frac{1}{1+\frac{\gamma^2}{|\alpha|^2}\frac{\bar n_{th}}{(gN_{pair})^2}}.
\end{eqnarray}

We make detail analysis of efficiency and other thermodynamic quantities in terms of thermal photon number and mean number of cavity photons by exploring various possible initialized states for experimental parameters of $\kappa/2\pi=74~\text{kHz}$ and $g/2\pi=334~\text{kHz}$ \cite{ar17,ar43}, and $\delta$ is controlled between $0$ and $-1~\text{MHz}$. We choose an interaction time $\tau=0.03/g$, which is a reasonable experimental parameter since the engine piston senses a strong radiation pressure with low thermal excitations (Appendix \ref{apa2}).

\begin{figure}[th]
\begin{center}
\hspace{-2.75cm}{(a)\hspace{3.5cm}(b)}\\
\includegraphics[width=.5\textwidth]{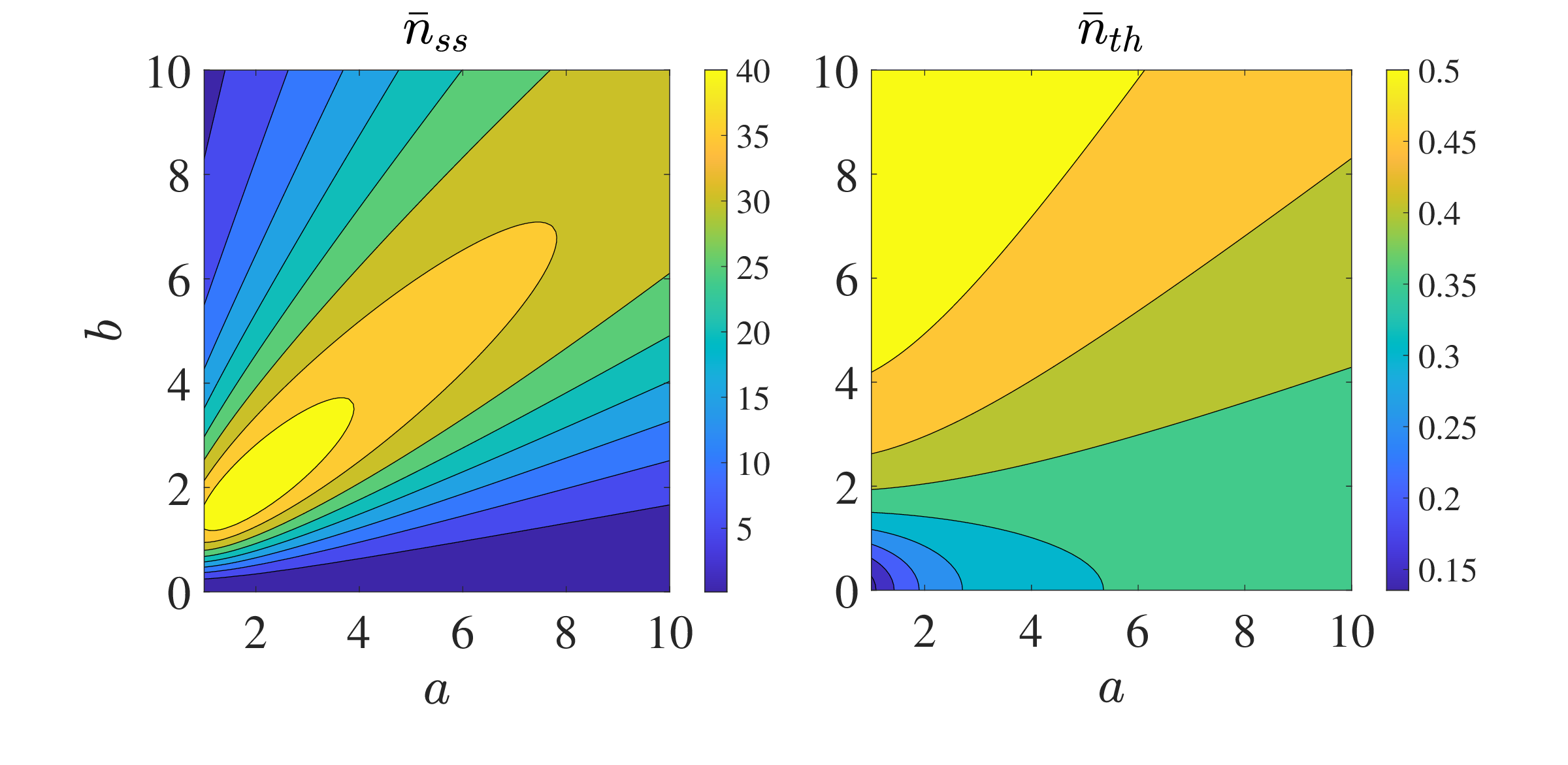}\\
\hspace{-2.75cm}{(c)\hspace{3.5cm}(d)}\\
\includegraphics[width=.5\textwidth]{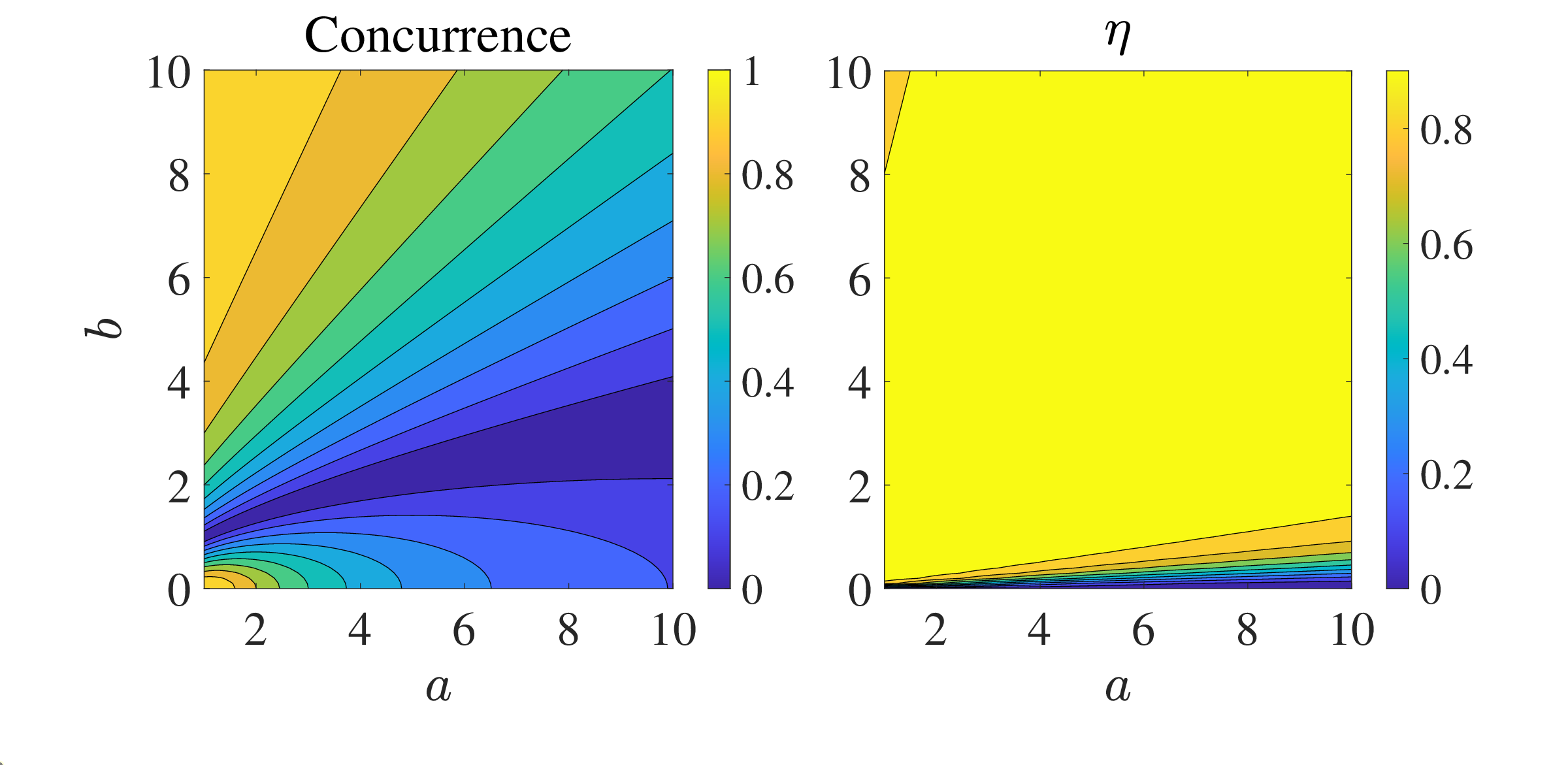}
 \caption{Effect of quantum-correlated atoms on engine performance. (a) Mean photon number of the cavity field, (b) thermal photon number, (c) concurrence, and (d) operating efficiency of the cavity field for the initialized state $|\psi\rangle={\cal N}\big(a|e_1e_2\rangle+b\big(|g_1e_2\rangle+e^{i\phi}|e_1g_2\rangle\big)+c|g_1g_2\rangle\big)$, $c=1, N_{pair}=2$ and $g\tau=0.03$. The other parameters are the same as in Fig. \ref{f2}.}\label{f3}
\end{center}
\end{figure}
\subsection{Engine performance analysis}
First, we investigate the most general case where $a,b,c\neq0.$ In Figs. \ref{f3}(a) and \ref{f3}(b), we vary $a$ and $b$ by fixing $c=1$, where we present the mean photon number, thermal photon number, concurrence, and analyze the performance of the engine by calculating the efficiency. The entanglement between two atoms can be quantified as concurrence in Fig. \ref{f3}(c), which is contributed by both the double excitations (Bell states with $b=0$) and singly-excited state components (Bell states with $b>>1$), except in the case when $a>>b$. Entanglement contribution in the singly-excited states rapidly depletes the mean photon number of the cavity field, but slowly increases the thermal photon number. The maximum values of $\bar{n}_{th}$ are reached where the concurrence is maximal, as can be seen in Figs. \ref{f3}(b) and \ref{f3}(c). In contrast, $\bar{n}_{ss}$ decreases, and its peak occurs for certain values of $a$ and $b$ that do not correspond to maximal entanglement. It is interesting to see that the efficiency is close to unity where the engine produces a large amount of work output, $W_{net}\propto\bar{n}_{ss}-\bar{n}_{th}$ (see Fig. \ref{f2}). Our photonic engine is fueled by quantum-correlated atoms that provide heat and work for the engine. This has increased the engine efficiency to almost unity, including a coherent state regime attained when $a=b=c$ as demonstrated in Fig. \ref{f3}(d). Engine efficiency and work output show the same trend, both being reduced, for large populations in the doubly-excited state and near maximal entanglement. The former case is obvious since coherence is lost for large $a$, which considerably lowers the radiation pressure, work output and engine efficiency. The latter case is due to the fact that maximally entangled states do not produce a temperature difference (heat gradient) between the effective hot and cold reservoirs. High heat gradient across reservoirs is associated with highly efficient work extraction, but this is not the case for maximally entangled atoms (see Sec. \ref{sec: III}).

\begin{figure}[b]
\begin{center}
\resizebox{0.5\textwidth}{!}{
\includegraphics{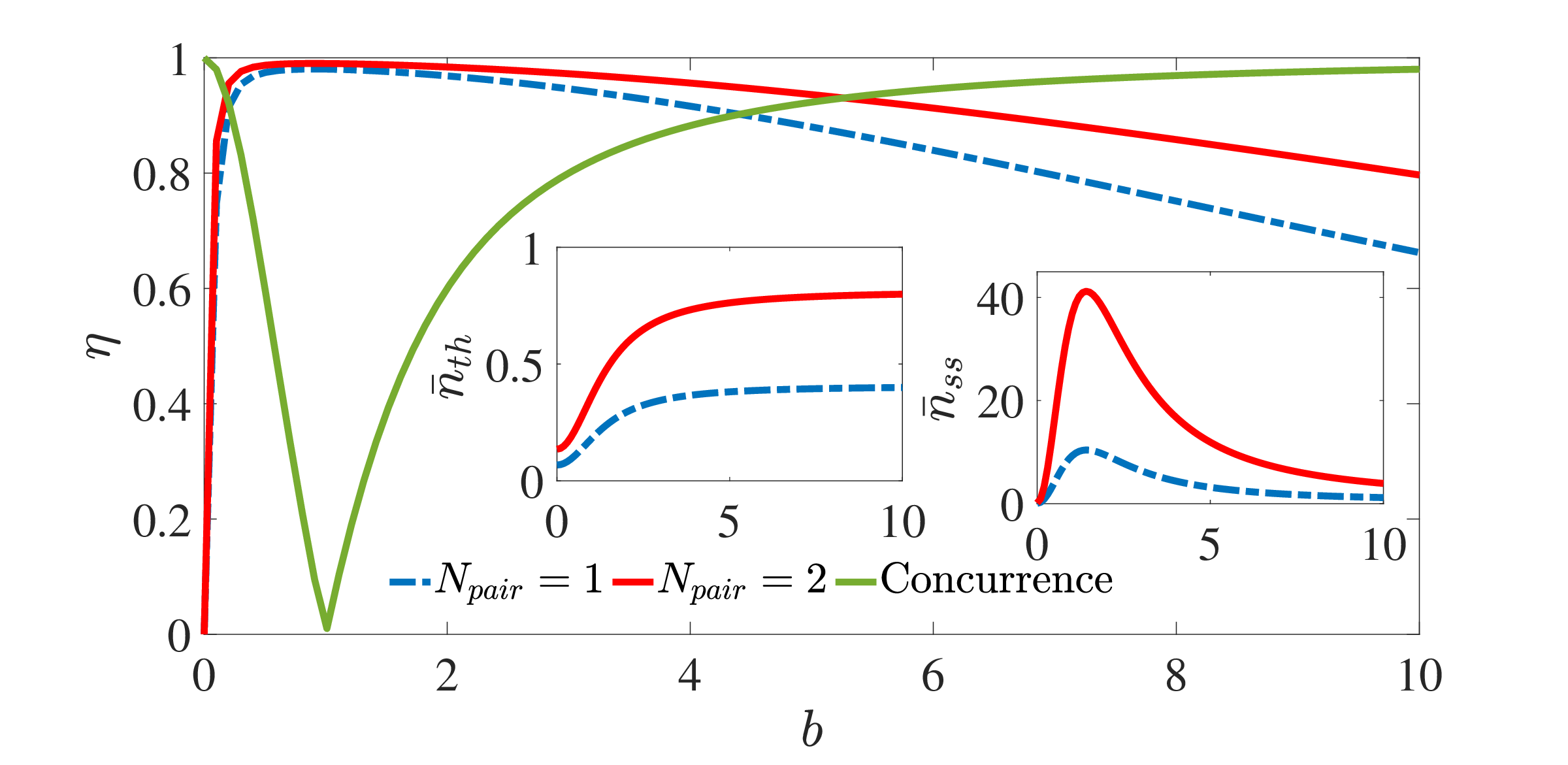}}
\caption{Operating efficiency of the cavity field, with mean photon number of the cavity ($\bar{n}_{ss}$) and thermal photon number ($\bar{n}_{th}$) displayed in the inset plots for $a=1$ and $c=1$. Concurrence is represented by a solid green curve, while other thermodynamic quantities are represented by dash-dot blue and solid red curves, respectively, for $N_{pair}=1$ and $N_{pair}=2$. The other parameters are the same as in Fig. \ref{f3}.}\label{f5}
\end{center}
\end{figure}
We explore the latter scenario in Fig. \ref{f5} by plotting the aforementioned quantities against $b$ to demonstrate how entanglement modifies the thermodynamic properties of the cavity field. Here we investigate engine's performance by operating the system between double excitations and singly-excited states, for which efficiency is totally lost. As we have mentioned earlier in Fig. \ref{f3}, the double excitations are related to the squeezing property of the cavity field and do not transport energy from the reservoir to the cavity photons - average number of photons in the cavity, work output and engine efficiency are negligible. However, singly-excited states, together with non-zero populations in the doubly-excited state and ground state, tends to raise the effective temperature of the reservoir more than that of the engine. The best operating efficiency occurs when the entanglement metric by concurrence vanishes. We note that when $a=b=c\neq0$, the state of the two-atoms is a tensor product of the maximum coherent superposition states of the individual atoms: $|\psi(\phi=0)\rangle=\big(|g_1\rangle+|e_1\rangle\big)/\sqrt{2}\otimes\big(|g_2\rangle+|e_2\rangle\big)/\sqrt{2}$. In this particular case, our photonic engine reduces to Ref. \cite{ar17} where it is fueled by atomic coherence rather than quantum entanglement. We emphasize here that high efficiency and large work output can be achieved as well for a finite entanglement. Realization of an arbitrary entangled state demonstrated in Figs. \ref{f3} and \ref{f5} should not be hard, for example, via single-qubit rotations and two-qubit gate operations. Another intriguing situation is when $a,c\neq0,b=0$, for which the normalized atomic state reduces to a Bell state, $|\psi(\phi=0)\rangle=\big(|g_1g_2\rangle+|e_1e_2\rangle\big)/\sqrt{2}$, which does not drive the cavity field and therefore the engine efficiency vanishes. The same holds for the other Bell states for $b>>a,c$ or $a,c=0,b\neq0$ in which our general state changes to $|\psi(\phi=0)\rangle=\big(|g_1e_2\rangle+|e_1g_2\rangle\big)/\sqrt{2}$. 

\begin{figure}
\begin{center}
\resizebox{0.5\textwidth}{!}{
\includegraphics{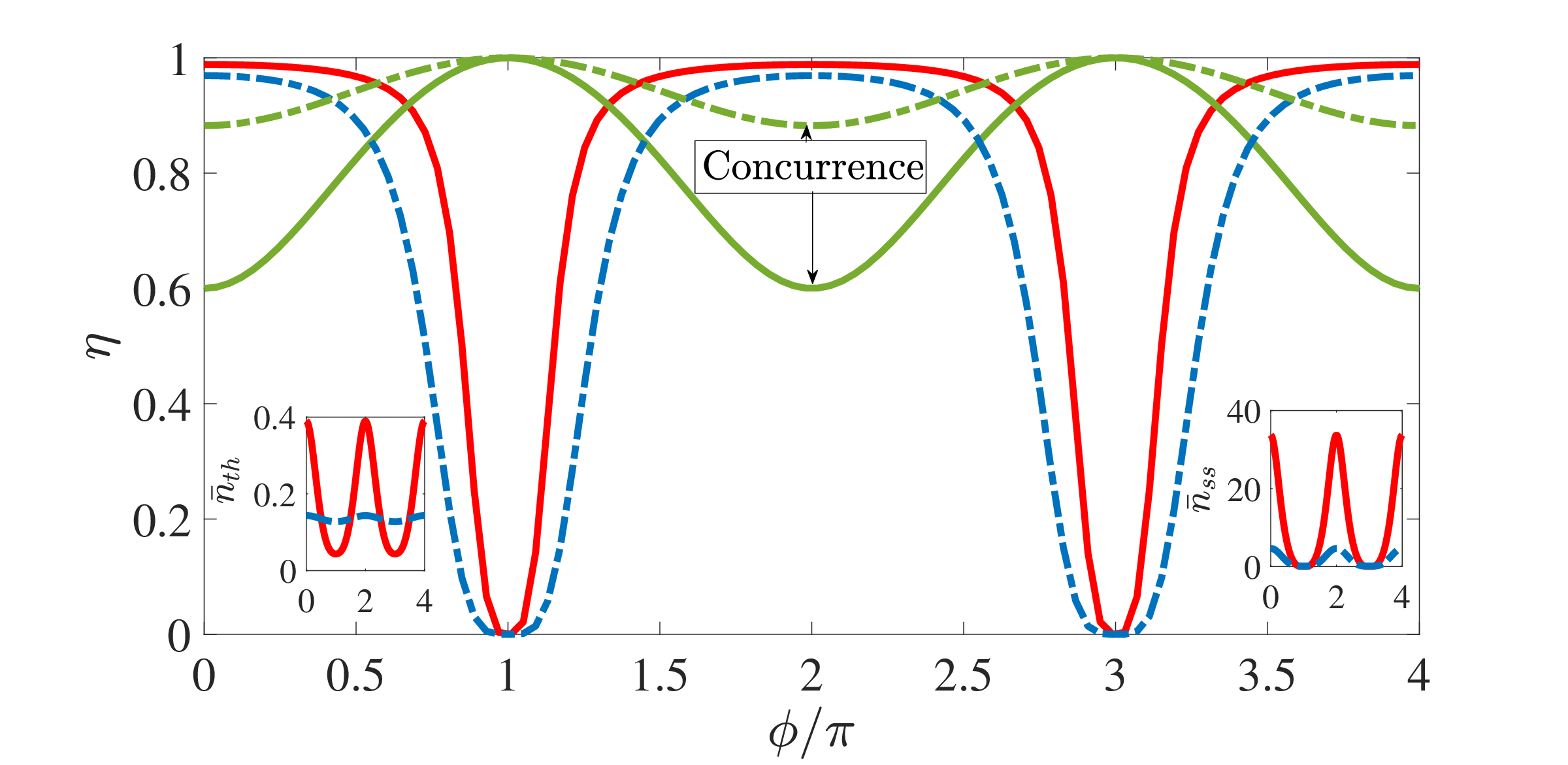}}
\caption{Operating efficiency of the cavity field, with mean photon number of the cavity ($\bar{n}_{ss}$) and thermal photon number ($\bar{n}_{th}$) displayed in the inset plots. Concurrence is represented by a dash-dot and solid green curves, while other thermodynamic quantities are represented by dash-dot blue and solid red curves for $b=0.25$ and $b=2$, respectively, with $a=1$. The other parameters are the same as in Fig. \ref{f3}.}\label{ff3}
\end{center}
\end{figure}
Moreover, in Fig. \ref{ff3} we study the effect of the relative phase between singly-excited states on the engine efficiency. We choose $b=0.25$ and $2$ as examples and vary $\phi\in[0,4\pi]$ to see how entanglement affects the efficiency and mean number of cavity photons. The odd $\phi=\pi(2s+1)$ and even $\phi=2\pi s$ with an integer $s$ indicate the subradiant and the superradiant states of the atomic pairs \cite{ar44}. The observed entanglement and efficiency oscillate in $\phi$, and high efficiency over a wide range of $\phi$ is shown for the superradiant state, as energy transfer from the effective reservoir to the cavity field is much enhanced due to constructive interference. In contrast, the efficiency is negligible for subradiant state owing to the destructive interference in an opposite energy transfer in the superradiant state. These quantum interferences arise from the fact that the single-mode cavity field cannot distinguish between pairs of atomic beams due to the symmetrical nature of the Jaynes-Cumming Hamiltonian \cite{ar23,ar45}.

\begin{figure}
\begin{center}
\hspace{-6.5cm}(a)\\
\includegraphics[width=.5\textwidth]{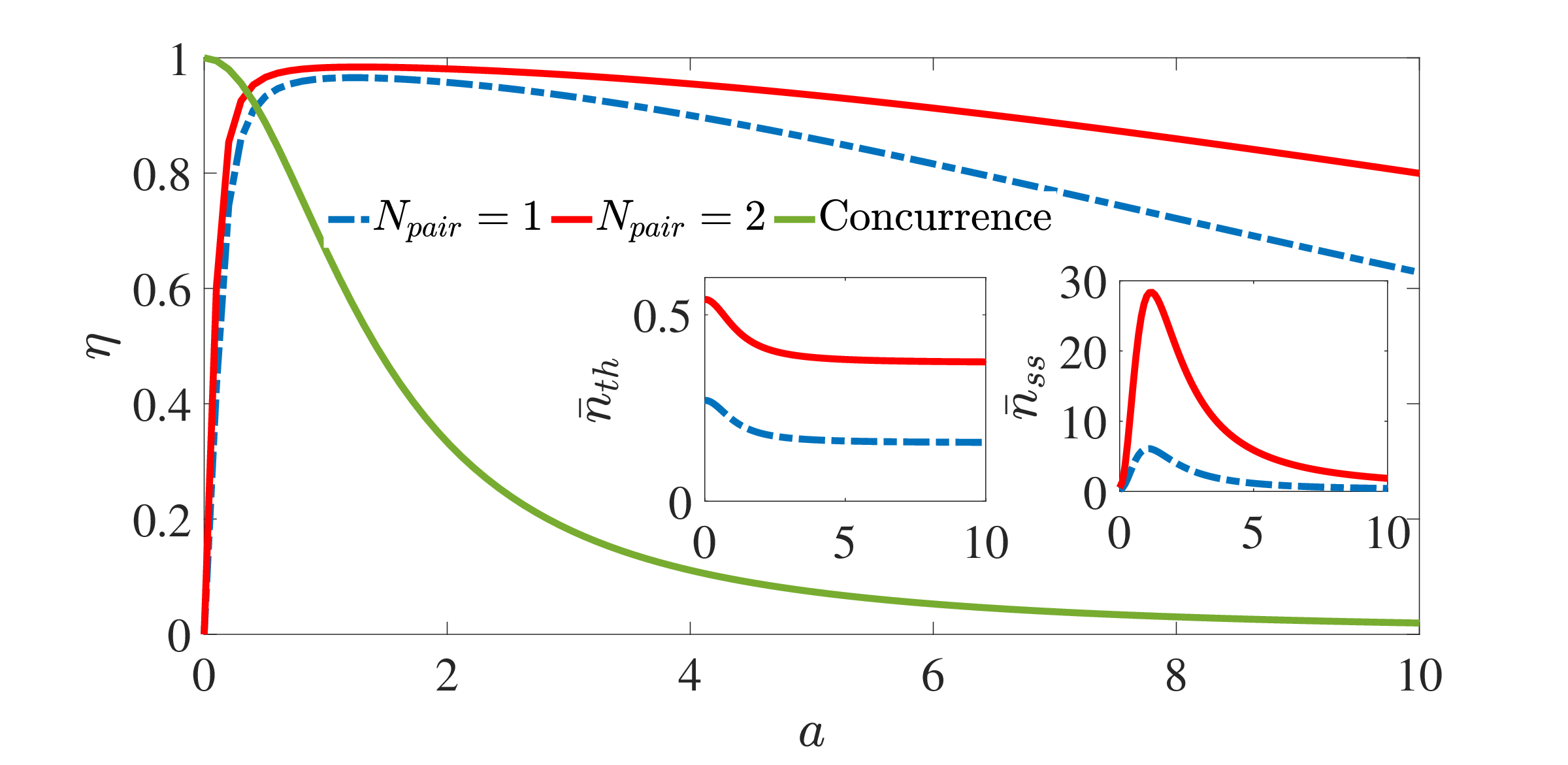}\\
\hspace{-6.5cm}(b)\\
\includegraphics[width=.5\textwidth]{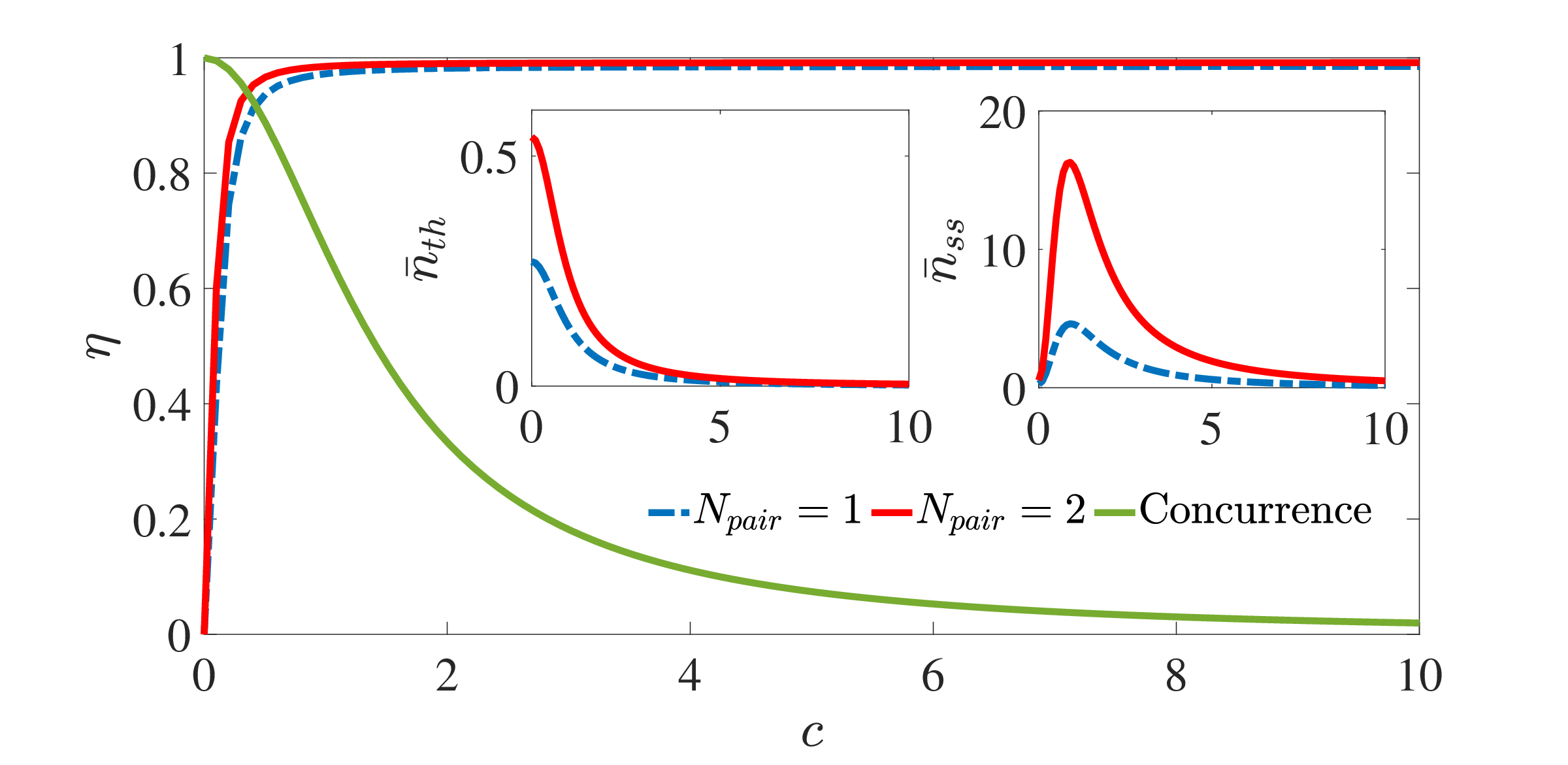}
\caption{Operating efficiency of the cavity field, with mean photon number of the cavity ($\bar{n}_{ss}$) and thermal photon number ($\bar{n}_{th}$) displayed in the inset plots, for (a) $b=1$ and $c=0$ and (b) $a=0$ and $b=1$. Concurrence is represented by a solid green curve, while other thermodynamic quantities are represented by dash-dot blue and solid red curves, respectively, for $N_{pair}=1$ and $N_{pair}=2$. The other parameters are the same as in Fig. \ref{f3}.}\label{f7}
\end{center}
\end{figure}
Next we focus on the double excitation states and study its effect of entanglement on the engine efficiency when $c=0$ in Fig. \ref{f7}. This represents a transient state of the system when depleted from the ground states. Photonic engine efficiency is low when $a$ is small, for which absorption of photons is more likely than emissions. Thus, the radiation pressure generated by the cavity photons is insignificant that it cannot produce an efficient work output. However, it is feasible to obtain high efficiency and substantial work output by increasing atomic populations in the doubly-excited state which facilitates emissions of more photons in the cavity. This obviously requires the presence of finite quantum correlations, as illustrated in Fig. \ref{f7}(a). Significant efficiency can be sustained within finite entanglement up to an $80\%$ of it, but the mean photon number rapidly drops as concurrence increases, so does the engine efficiency. We can also see that, if quantum correlations are minimal, populating the doubly-excited state ($a>>1$) alone is not much helpful for the engine performance.

Finally, we investigate a weak excitation regime where the double excitation state is negligible.
This case is depicted in Fig. \ref{f7}(b), where the pairs of atoms introduced into the cavity mainly serve as the engine's driving force without causing significant excitations in the system, in particular for $c>>1$. The effective excited population as well as the thermal photon number are hence negligibly small, which causes the engine efficiency to increase to almost unity. However, the work output is not promising since radiation pressure represented by the average number of photons falls there. Negligible work output with high efficiency is practically not useful. We can still achieve a considerable amount of work output, though not as high as demonstrated before, with finite quantum correlations resulting from singly-excited states when the double ground state populations are not dominant ($c\sim b$).
\section{\label{sec: V} conclusion}
We investigated a photonic engine powered by an effective reservoir consisting pairs of two-level atoms and a vacuum environment. The reservoir contains quantum correlations that have been extensively studied as an a resource for quantum information processing applications. Here we have theoretically studied the role of two-atom entanglement in effective work extraction processes by means of a photonic engine operating with an analogous classical string cycle. The engine cycles between different states of the effective reservoir acting as a hot reservoir for finite driving $\alpha$ and a cold reservoir for the non-driven regime. Transfer of energy from the effective reservoir to the engine modifies the thermodynamic properties of the cavity field. Several overlapping regimes of high efficiency and large work output have been observed. The engine works well by exciting atoms with a finite entanglement or coherence between them because this promotes an effective re-emission of photons into the cavity and this in turn increases the radiation pressure and amount of work output. Engine efficiency has increased to near unity due to significant transfer of energy from the effective reservoir to the engine \cite{ar39,ar40}. This high efficiency is observed when the state of the two atoms is a tensor product of the maximally coherent superposition states of the individual atoms consistent with Ref. \cite{ar17}. Engine efficiency close to unity can also be achieved for finite entanglement provided proper selections of phase and atomic populations are made. Realization of these an arbitrary entangled states should not be hard, for example, via single-qubit rotations and two-qubit gate operations. In contrast, the effective reservoir serves as a passive state to the cavity field for maximally entangled Bell states. In this case, constantly sending quantum-correlated atoms into the cavity does not create a heat gradient by driving the cavity field from the effective cold reservoir to hot reservoir. As a result, the engine is inefficient and produces no work output because the cavity field maintains thermal equilibrium with the effective cold reservoir. 

In this work, we have investigated the influence of coherent and entangled reservoirs on the operation of a quantum engine with a cavity field as its working medium. Future contributions in photonic engines may involve non-thermal reservoirs made up of many-body systems with collective effects as well as additional quantum correlations including discord, steering, and Bell non-locality \cite{ar49}. We have also chosen work output and efficiency as the engine's performance metrics, while one could also test other newly developing quality and quantity measures like reliability \cite{co3}, which considers work and heat as fluctuating quantities, and time required for thermodynamic processes.
\begin{acknowledgments}
We acknowledge support from the National Science and Technology Council (NSTC), Taiwan, under Grants No. 112-2112-M-001-079-MY3, and No. NSTC-112-2119-M-001-007. We are also grateful for support from TG 1.2 of NCTS at NTU.
\end{acknowledgments}
\appendix
\section{\label{apa}The superoperator and master equation}
In the micromaser model, light matter interaction is a random process, where the chance for the light-atom interaction to occur in a time interval $t+\Delta t$ is $r_a\Delta t$, in which $r_a$ is an atomic fly rate. It is thus obvious that there will be no interaction in the interval $1-r_a\Delta t$. The total state of the cavity field is the sum of the two possible outcomes put in the form \cite{ar31,ar32}
\begin{eqnarray*}
 \rho(t+\Delta t)=r_a\Delta t{\rm Tr_a}[U(\tau)\rho_{ac}(t)U^\dag(\tau)]+(1-r_a\Delta t)\rho(t).
\end{eqnarray*}
In the limit $\Delta t\rightarrow0$, the density matrix of the cavity system can be written as:
\begin{eqnarray}
 \frac{d\rho(t)}{dt}&=&r_a[S(\tau)-1]\rho(t).\label{10.0}
\end{eqnarray}
where $S(\tau)\rho(t)={\rm Tr_a}[U(\tau)\rho_{ac}(t)U^\dag(\tau)]$ is a superoperator acting on the density operator to evolve the state of the cavity field from $\rho(t_i)$ to $\rho(t_i+\tau)$.

The unitary time evolution operator $U(\tau)$ corresponding to $\hat H_I$ is determined by treating the two atoms as a non-interacting spin $1/2$ particles added to give a triplet state with $U_1(\tau)$ and a singlet state with $U_0(\tau)$: $U(\tau)=U_1(\tau)\oplus U_0(\tau)=U_{1/2}(\tau)\otimes U_{1/2}(\tau)$, where $U_{1/2}(\tau)$ is a unitary operator for the single two-level atom \cite{ar31,ar33}. The unitary evolution operator for the two atoms, spanned by a natural basis $\{|1\rangle\equiv|e_1e_2\rangle,|2\rangle\equiv|e_1g_2\rangle,|3\rangle\equiv|g_1e_2\rangle,|4\rangle\equiv|g_1g_2\rangle\}$, can be written by setting $\hbar=1$ as
\begin{eqnarray*}
U(\tau)=e^{-i\hat H_I\tau}=\begin{pmatrix}
    u_{11}&u_{12}&u_{13}&u_{14}\\
   u_{21}&u_{22}&u_{23}&u_{24}\\
  u_{31}&u_{32}&u_{33}&u_{34}\\
  u_{41}&u_{42}&u_{43}&u_{44}\\
   \end{pmatrix}\label{u1},
\end{eqnarray*}
where 
\begin{eqnarray*}
u_{11}&=&[\text{cos}(\frac{\Omega_{1,n}\tau}{2})-\frac{i\delta_1\tau}{2}\text{sinc}(\frac{\Omega_{1,n}\tau}{2})][\text{cos}(\frac{\Omega_{2,n}\tau}{2})\nonumber\\&-&\frac{i\delta_2\tau}{2}\text{sinc}(\frac{\Omega_{2,n}\tau}{2})],\\
u_{12}&=&-ig\tau[\text{cos}(\frac{\Omega_{1,n}\tau}{2})-\frac{i\delta_1\tau}{2}\text{sinc}(\frac{\Omega_{1,n}\tau}{2})]\nonumber\\&\times&\text{sinc}(\frac{\Omega_{2,n}\tau}{2})\hat a,\\
u_{13}&=&-ig\tau\text{sinc}(\frac{\Omega_{1,n}\tau}{2})\hat a[\text{cos}(\Omega_{2,n}\tau/2)-\frac{i\delta_2\tau}{2}\nonumber\\&\times&\text{sinc}(\frac{\Omega_{2,n}\tau}{2})],\\
\hspace{-1.5cm}u_{14}&=&-(g\tau)^2\text{sinc}(\frac{\Omega_{1,n}\tau}{2})\hat a[\text{sinc}(\frac{\Omega_{2,n}\tau}{2})]\hat a,\\
u_{21}&=&-ig\tau[\text{cos}(\frac{\Omega_{1,n}\tau}{2})-\frac{i\delta_1\tau}{2}\text{sinc}(\frac{\Omega_{1,n}\tau}{2})]\hat a^\dag\nonumber\\&\times&\text{sinc}(\frac{\Omega_{2,n-1}\tau}{2}),\\
u_{22}&=&[\text{cos}(\frac{\Omega_{1,n}\tau}{2})-\frac{i\delta_1\tau}{2}\text{sinc}(\frac{\Omega_{1,n}\tau}{2})][\text{cos}(\frac{\Omega_{2,n-1}\tau}{2})\nonumber\\&+&\frac{i\delta_2\tau}{2}\text{sinc}(\frac{\Omega_{2,n-1}\tau}{2})],\\
u_{23}&=&-(g\tau)^2\text{sinc}(\frac{\Omega_{1,n}\tau}{2})\hat a\hat a^\dag[\text{sinc}(\frac{\Omega_{2,n-1}\tau}{2})],\\
u_{24}&=&-ig\tau\text{sinc}(\frac{\Omega_{1,n}\tau}{2})\hat a[\text{cos}(\frac{\Omega_{2,n-1}\tau}{2})+\frac{i\delta_2\tau}{2}\nonumber\\&\times&\text{sinc}(\frac{\Omega_{2,n-1}\tau}{2})],\\
u_{31}&=&-ig\tau\hat a^\dag\text{sinc}(\frac{\Omega_{1,n-1}\tau}{2})[\text{cos}(\frac{\Omega_{2,n}\tau}{2})-\frac{i\delta_2\tau}{2}\nonumber\\&\times&\text{sinc}(\frac{\Omega_{2,n}\tau}{2})],\\
u_{32}&=&-(g\tau)^2\hat a^\dag\text{sinc}(\frac{\Omega_{1,n-1}\tau}{2})[\text{sinc}(\frac{\Omega_{2,n}\tau}{2})]\hat a,\\
u_{33}&=&[\text{cos}(\frac{\Omega_{1,n-1}\tau}{2})+\frac{i\delta_1\tau}{2}\text{sinc}(\frac{\Omega_{1,n-1}\tau}{2})][\text{cos}(\frac{\Omega_{2,n}\tau}{2})\nonumber\\&-&\frac{i\delta_2\tau}{2}\text{sinc}(\frac{\Omega_{2,n}\tau}{2})],\\
u_{34}&=&-ig\tau[\text{cos}(\frac{\Omega_{1,n-1}\tau}{2})+\frac{i\delta_1\tau}{2}\text{sinc}(\frac{\Omega_{1,n-1}\tau}{2})]\nonumber\\&\times&\text{sinc}(\frac{\Omega_{2,n}\tau}{2})\hat a,\\
u_{41}&=&-(g\tau)^2\hat a^\dag\text{sinc}(\frac{\Omega_{1,n-1}\tau}{2})\hat a^\dag[\text{sinc}(\frac{\Omega_{2,n-1}\tau}{2})],\\
u_{42}&=&-ig\tau\hat a^\dag\text{sinc}(\frac{\Omega_{1,n-1}\tau}{2})[\text{cos}(\frac{\Omega_{2,n-1}\tau}{2})\nonumber\\&+&\frac{i\delta_2\tau}{2}\text{sinc}(\frac{\Omega_{2,n-1}\tau}{2})],
\end{eqnarray*}
\begin{eqnarray*}
u_{43}&=&-ig\tau[\text{cos}(\frac{\Omega_{1,n-1}\tau}{2})+\frac{i\delta_1\tau}{2}\text{sinc}(\frac{\Omega_{1,n-1}\tau}{2})]\hat a^\dag\nonumber\\&\times&\text{sinc}(\frac{\Omega_{2,n-1}\tau}{2}),\\
u_{44}&=&[\text{cos}(\frac{\Omega_{1,n-1}\tau}{2})+\frac{i\delta_1\tau}{2}\text{sinc}(\frac{\Omega_{1,n-1}\tau}{2})]\nonumber\\&\times&[\text{cos}(\frac{\Omega_{2,n-1}\tau}{2})+\frac{i\delta_2\tau}{2}\text{sinc}(\frac{\Omega_{2,n-1}\tau}{2})],
\end{eqnarray*}
with $\Omega_{1(2),n}=\sqrt{\delta^2_{1(2)}+4g^2(\hat a^\dag \hat a+1)}$ and $\Omega_{1(2),n-1}=\sqrt{\delta^2_{1(2)}+4g^2\hat a^\dag \hat a}$ representing the generalized Rabi frequencies.

The superoperator can be expanded in terms of the matrix elements of atomic density matrix and unitary evolution operator as:  $S(\tau)\rho(t)={\rm Tr_a}[U(\tau)\rho_{ac}(t)U^\dag(\tau)]=\sum^4_{i,j=1}\rho_{ij}\sum^4_{k}u_{ki}\rho(t)u^\dag_{kj}.$
We use the bosonic commutation relation $[\hat a, \hat a^\dag]=1$ to explicitly calculate the superoperator in the Markovian regime ($g\tau<<1$). In this regime, the trigonometric functions involved in the unitary evolution operator can be approximated as: $\text{cos}(\frac{\Omega_{1(2),n}\tau}{2})\approx\text{cos}(\frac{\delta_{1(2)}\tau}{2})-\frac{1}{2}\text{sinc}(\frac{\delta_{1(2)}\tau}{2})(g\tau)^2(\hat a^\dag \hat a+1),$ and $\text{sinc}(\frac{\Omega_{1(2),n}\tau}{2})\approx\text{sinc}(\frac{\delta_{1(2)}\tau}{2})+\frac{2}{(\tau\delta_{1(2)})^2}\{\text{cos}(\frac{\delta_{1(2)}\tau}{2})-\text{sinc}(\frac{\delta_{1(2)}\tau}{2})\}(g\tau)^2(\hat a^\dag\hat a+1)$. The superoperator contains 64 terms and they can be obtained by straightforward calculations. The diagonal terms can be written up to second order in $g\tau$ as
\begin{eqnarray*}
u_{11}\rho(t)u^\dag_{11}&=&\rho(t)-(g\tau)^2[R^\ast\rho(t)\hat a \hat a^\dag+R\hat a \hat a^\dag\rho(t)],\\
u_{22}\rho(t)u^{\dag}_{22}&=&\rho(t)-(g\tau)^2 [E\rho(t)\hat{a}\hat{a}^{\dag}+D^{\ast}\rho(t)\hat{a}^{\dag}\hat{a}\nonumber\\&+&E^{\ast}\hat{a}^{\dag} \hat{a}\rho(t)+D\hat{a}\hat{a}^{\dag}\rho(t)],\\
u_{33}\rho(t)u^{\dag}_{33}&=&\rho(t)-(g\tau)^2 [E^{\ast}\rho(t)\hat a^{\dag}\hat a+D\rho(t)\hat a \hat a^\dag\nonumber\\&+&E\hat a\hat a^\dag \rho(t)+D^{\ast}\hat a^{\dag} \hat{a}\rho(t)],\\
u_{44}\rho(t)u^\dag_{44}&=&\rho(t)-(g\tau)^2[R^\ast\hat a \hat a^\dag\rho(t)+R\rho(t)\hat a \hat a^\dag],
\end{eqnarray*}
where $R=D+E$, $D=a_1e^{i\delta_1\tau/2}+ib_1e^{i\delta_1\tau/2}$, $E=a_2e^{i\delta_2\tau/2}+ib_2e^{i\delta_2\tau/2}$, $a_{1(2)}=\frac{1}{2}\text{sinc}(\frac{\delta_{1(2)}\tau}{2})$ and $b_{1(2)}=\frac{1}{\delta_{1(2)}\tau}[\text{cos}(\frac{\delta_{1(2)}\tau}{2})-\text{sinc}(\frac{\delta_{1(2)}\tau}{2})]$.
After carrying out all the calculations, one can get a full expression of the superoperator written as
\begin{widetext}
\begin{eqnarray}
 S(\tau)\rho(t)&=&\rho_{11}\rho(t)-\rho_{11}(g\tau)^2\bigg((R^\ast\rho(t)\hat a \hat a^\dag+R\hat a \hat a^\dag\rho(t)-(\text{sinc}^2\frac{\delta_1\tau}{2}+\text{sinc}^2\frac{\delta_2\tau}{2})\hat a^\dag\rho(t)\hat a\bigg)+\rho_{22}\rho(t)-(g\tau)^2\rho_{22}\nonumber\\&\times&\bigg(E\rho(t)\hat{a}\hat{a}^{\dag}+D^{\ast}\rho(t)\hat{a}^{\dag}\hat{a}+E^{\ast}\hat{a}\hat{a}^{\dag}\rho(t)+D\hat{a}^{\dag}\hat{a}\rho(t)-(\text{sinc}^2\frac{\delta_1\tau}{2}\hat a^\dag\rho(t)\hat a+\text{sinc}^2\frac{\delta_2\tau}{2}\hat a\rho(t)\hat a^\dag)\bigg)\nonumber\\&+&\rho_{33}\rho(t)-(g\tau)^2\rho_{33}\bigg(E^{\ast}\rho(t)\hat a^{\dag}\hat a+D\rho(t)\hat a \hat a^\dag+E\hat a^\dag\hat a \rho(t)+D^{\ast}\hat a\hat{a}^{\dag}\rho(t)-(\text{sinc}^2\frac{\delta_1\tau}{2}\hat a\rho(t)\hat a^\dag)\nonumber\\&+&\text{sinc}^2\frac{\delta_2\tau}{2}\hat a^\dag\rho(t)\hat a\bigg)+\rho_{44}\rho(t)-\rho_{44}(g\tau)^2\bigg(R^\ast\hat a \hat a^\dag\rho(t)+R\rho(t)\hat a \hat a^\dag-(\text{sinc}^2\frac{\delta_1\tau}{2}+\text{sinc}^2\frac{\delta_2\tau}{2})\hat a\rho(t)\hat a^\dag\bigg)\nonumber\\&+&\rho_{12}\bigg(ig\tau\text{sinc}\frac{\delta_2\tau}{2}e^{-i\delta_2\tau/2}(\rho(t)\hat a^\dag-\hat a^\dag\rho(t))\bigg)+\rho_{13}\bigg(ig\tau\text{sinc}\frac{\delta_1\tau}{2}e^{-i\delta_1\tau/2}(\rho(t)\hat a^\dag-\hat a^\dag\rho(t))\bigg)\nonumber\\&+&\rho_{14}\bigg((g\tau)^2\text{sinc}\frac{\delta_1\tau}{2}\text{sinc}\frac{\delta_2\tau}{2}e^{-i(\delta_1+\delta_2)\tau/2}(2\hat a^\dag\rho(t)\hat a^\dag-\hat a^{\dag2}\rho(t)-\rho(t)\hat a^{\dag2})\bigg)+\rho_{21}\bigg(ig\tau\text{sinc}\frac{\delta_2\tau}{2}e^{i\delta_2\tau/2}\nonumber\\&\times&(\rho(t)\hat a-\hat a\rho(t))\bigg)+\rho_{23}\bigg((g\tau)^2\text{sinc}\frac{\delta_1\tau}{2}\text{sinc}\frac{\delta_2\tau}{2}e^{-i(\delta_1-\delta_2)\tau/2}(\hat a^\dag\rho(t)\hat a+\hat a\rho(t)\hat a^\dag-\rho(t)\hat a\hat a^{\dag}-\hat a^\dag\hat a\rho(t))\bigg)\nonumber\\&+&\rho_{24}\bigg(ig\tau\text{sinc}\frac{\delta_1\tau}{2}e^{-i\delta_1\tau/2}(\rho(t)\hat a^\dag-\hat a^\dag\rho(t))\bigg)+\rho_{31}\bigg(ig\tau\text{sinc}\frac{\delta_1\tau}{2}e^{i\delta_1\tau/2}(\rho(t)\hat a-\hat a\rho(t))\bigg)+\rho_{32}\bigg((g\tau)^2\nonumber\\&\times&\text{sinc}\frac{\delta_1\tau}{2}\text{sinc}\frac{\delta_2\tau}{2}e^{i(\delta_1-\delta_2)\tau/2}(\hat a^\dag\rho(t)\hat a+\hat a\rho(t)\hat a^\dag-\hat a^\dag\hat a\rho(t)-\rho(t)\hat a\hat a^\dag)\bigg)+\rho_{34}\bigg(ig\tau\text{sinc}\frac{\delta_2\tau}{2}e^{-i\delta_2\tau/2}\nonumber\\&\times&(\rho(t)\hat a^\dag-\hat a^\dag\rho(t))\bigg)+\rho_{41}\bigg((g\tau)^2\text{sinc}\frac{\delta_1\tau}{2}\text{sinc}\frac{\delta_2\tau}{2}e^{i(\delta_1+\delta_2)\tau/2}(2\hat a\rho(t)\hat a-\hat a^{2}\rho(t)-\rho(t)\hat a^{2})\bigg)\nonumber\\&+&\rho_{42}\bigg(ig\tau\text{sinc}\frac{\delta_1\tau}{2}e^{i\delta_1\tau/2}(\rho(t)\hat a-\hat a\rho(t))\bigg)+\rho_{43}\bigg(ig\tau\text{sinc}\frac{\delta_2\tau}{2}e^{i\delta_2\tau/2}(\rho(t)\hat a-\hat a\rho(t))\bigg).\label{so}
\end{eqnarray}
\end{widetext}
Master equation of the cavity photons described in Eq. (\ref{10.0}) can be rearranged into Lindblad forms \cite{ar46} by applying the normalization condition for the initial atomic states ($\rho_{11}+\rho_{22}+\rho_{33}+\rho_{44}=1$) and setting $\delta_1=\delta_2=\delta$. The master equation then becomes
\begin{eqnarray}
\frac{d\rho(t)}{dt}&=&-i[H_{\rm eff},\rho(t)]+p_1[2\hat a^\dag\rho(t)\hat a-(\hat a\hat a^\dag\rho(t)\nonumber\\&+&\rho\hat a\hat a^\dag)]+p_2[2\hat a\rho(t)\hat a^\dag-(a^\dag\hat a\hat\rho+\rho\hat a^\dag\hat a)]\nonumber\\&+&\mu(2\hat a\rho(t)\hat a-\hat a^{2}\rho(t)-\rho(t)\hat a^{2})\nonumber\\&+&\mu^\ast(2\hat a^\dag\rho(t)\hat a^\dag-\hat a^{\dag2}\rho(t)-\rho(t)\hat a^{\dag2}),\label{cm}
\end{eqnarray}
with corresponding effective Hamiltonian $H_{\rm eff}$, and coefficients $p_1$, $p_2$ and $\mu$ obtained in the main text.

We considered the interaction of the cavity photon with an external environment that contains infinite degrees of freedom. It is common to model the reservoir by a set of infinitely many harmonic oscillators. A general class of reservoir is squeezed thermal reservoir \cite{ar34}. Heat engine operation coupled to this reservoir has been realized in optomechanical oscillator and microwave experiments \cite{ar47, ar48}. In this work, an effective reservoir is engineered for the quantum system from pairs of atomic beams and vacuum environment represented by a dissipation $\frac{d\rho(t)}{dt}=\frac{\kappa}{2}(2\hat a\rho(t)\hat a^\dag-\rho(t)\hat a^\dag\hat a-\hat a^\dag\hat a\rho(t))$\cite{ar25}. We can define the mean excitation number $\bar n_{th}$ of the new reservoir provided that $\frac{\bar n_{th}}{\bar n_{th}+1}=\frac{p_1}{p_2+\frac{\kappa}{2}}$, which gives $\bar n_{th}=\frac{p_1}{\kappa/2+(p_2-p_1)}$. The temperature $T_{R}$ of the effective reservoir can be obtained by taking the ratio of the coefficients of the Lindblad operators corresponding to incoherent excitations and de-excitations and associating it with a Boltzmann factor. It is given by $T_{R}=\frac{\hbar\omega_a}{k_B\text{log}(1+\frac{1}{\bar n_{th}})},$ where $\omega_a=\frac{\omega_1+\omega_2}{2}$ and $k_B$ is the Boltzmann constant.
\section{\label{apa2} Mean number of cavity photons} 
\begin{figure}
\begin{center}
\resizebox{0.45\textwidth}{!}{\includegraphics{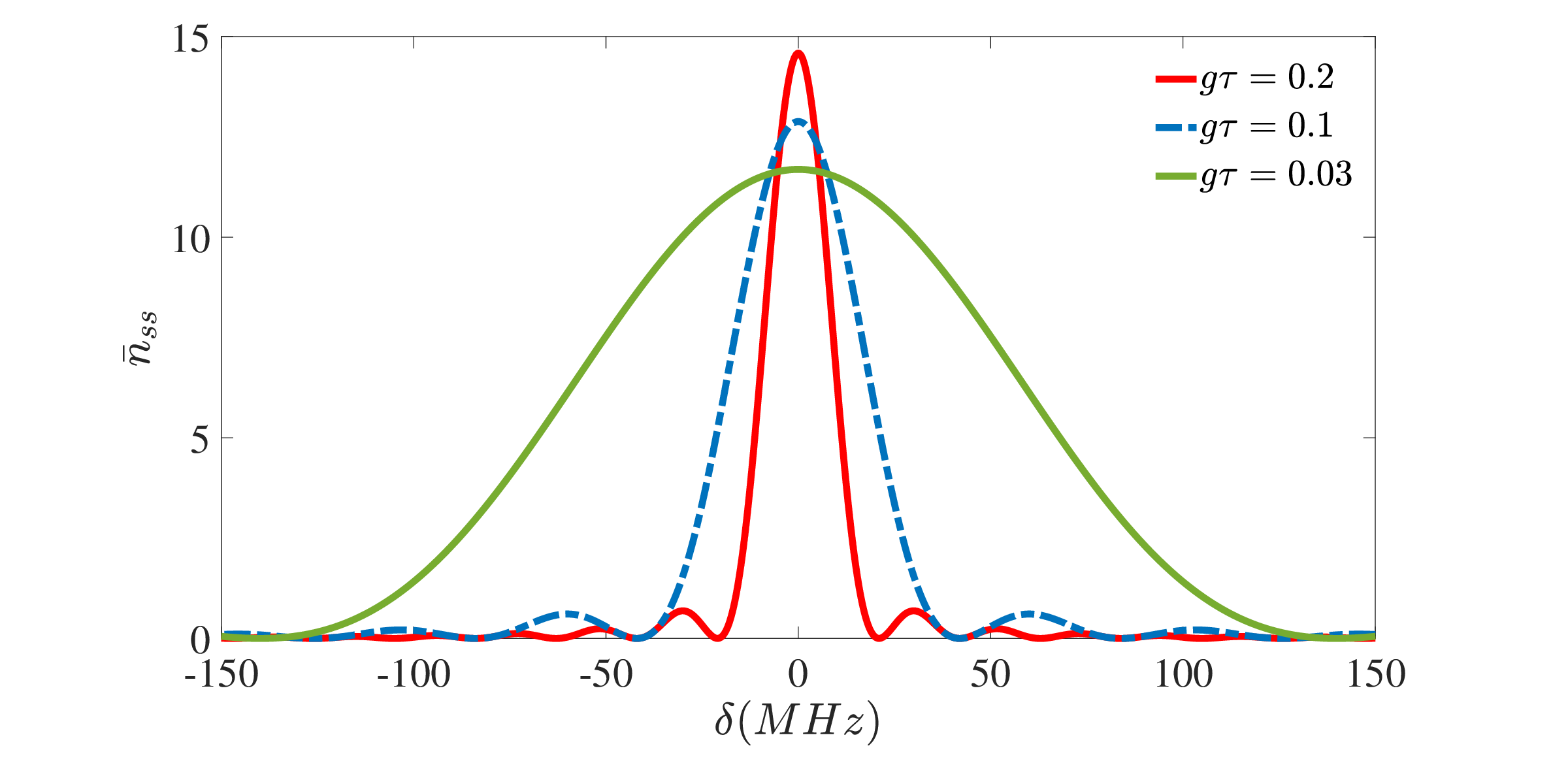}}
\resizebox{0.45\textwidth}{!}{\includegraphics{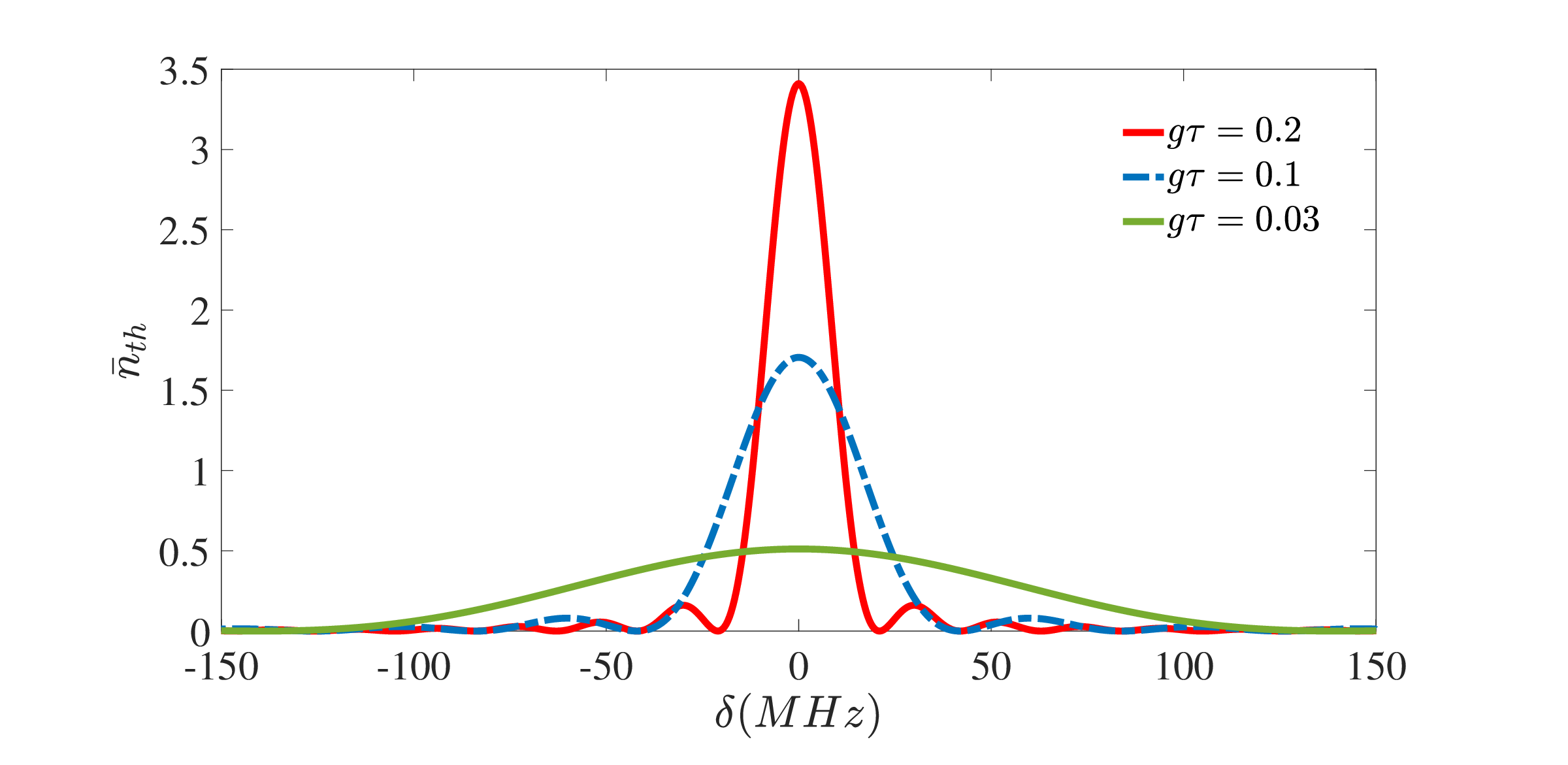}}
\caption{Mean photon number of the cavity field $\bar{n}_{ss}$ and thermal photon number ($\bar{n}_{th}$). The parameters are the same as in Fig. \ref{f2}. The engine piston experiences a strong radiation pressure in a relatively short interaction time $\tau=0.03/g$ because the average number of photons in the cavity is significant over a wide range of detuning frequency and at the same time the thermal photon number is low.}\label{f9}
\end{center}
\end{figure}
The time derivative of the mean photon number of the cavity is given by $\big\langle\frac{d\hat a^\dag\hat a}{dt}\big\rangle={\rm Tr}\big(\frac{d\rho(t)}{dt}\hat a^\dag\hat a\big).$ Using Eq. (\ref{cm2}) in the main paper, we get
\begin{eqnarray}
\left\langle\frac{d\hat a^\dag\hat a}{dt}\right\rangle&=&-2\gamma\langle\hat a^\dag\hat a\rangle-igN_{pair}(\alpha\langle\hat a^\dag\rangle\nonumber\\&-&\alpha^\ast\langle\hat a\rangle)+2\gamma\bar n_{th}.\label{mp2}
\end{eqnarray}
We clearly see that mean number of cavity photons has a well defined solution when $\gamma>0$. Since $\gamma=(p_2-p_1)+\kappa/2$, a regime where our system can operate is confined to $p_1<(p_2+\kappa/2)$ with a threshold condition $p_1=(p_2+\kappa/2)$.
 
To find a steady-state solution of Eq. (\ref{mp2}), we need to evaluate the time evolution of $\hat a$ and $\hat a^\dag$. Equation of motion for the annihilation operator is $\langle\frac{d\hat a}{dt}\rangle=-igN_{pair}\alpha-\gamma\langle\hat a\rangle$, and its Hermitian conjugate gives $\langle\frac{d\hat a^\dag}{dt}\rangle$. Steady-state mean photon number and thermal photon number are shown in Fig. \ref{f9} for different atom-field interaction times. The mean photon number of the cavity filed is marginally suppressed in a relatively short interaction time, while the thermal photon number is significantly reduced. It is possible to run the engine cycle through a broad range of frequencies in such a short interaction time. However, as the time of the interaction is increased, thermal photon number, which is a prominent source of fluctuation in experiments \cite{ar17}, increases especially near the resonance case. Therefore, a relatively short-interaction time is more feasible in experiments.

\nocite{*}

\bibliography{apssamp}% Produces the bibliography via BibTeX.

\end{document}